\documentclass{article}
\usepackage{geometry}                % See geometry.pdf to learn the layout options. There are lots.
\usepackage{graphicx}
\usepackage{amssymb}
\usepackage{epstopdf}
\usepackage{latexsym}
\usepackage{dcolumn}
\usepackage{epsfig}
\usepackage{epsfig}
\newcommand{\ba}{\begin{eqnarray}}

\newcommand{\ea}{\end{eqnarray}}
\newcommand{\be}{\begin{equation}}
\newcommand{\ee}{\end{equation}}
\newcommand{\beq}{\begin{equation}}
\newcommand{\eeq}{\end{equation}}
\newcommand{\beqarray}{\begin{eqnarray}}
\newcommand{\eeqarray}{\end{eqnarray}}
\newcommand{\bdisplay}{\begin{displaymath}}
\newcommand{\edisplay}{\end{displaymath}}

\title{Comment on ``Ultrahigh-Energy Neutrino-Nucleon Deep-Inelastic Scattering and the Froissart Bound'': Phys. Rev. Lett. {\bf 106}, 231802 (2011) }
\author{{Martin~M.~Block}\\
 {\it Department of Physics and Astronomy, Northwestern University, Evanston, IL 60208}
\and
Phuoc Ha\\
{\it Department of Physics, Astronomy and Geosciences, Towson University, Towson, MD 21252}
\and
Douglas W. McKay\\
 {\it Department of Physics and Astronomy, University of Kansas, Lawrence, KS 66045}}
\date{\today}
\begin{document}
\maketitle
\begin {abstract}
 The authors of a recent  paper, ``Ultrahigh-Energy Neutrino-Nucleon Deep-Inelastic Scattering and the Froissart Bound", A. Illarianov, B. Kniehl and A. Kotikov, Phys. Rev. Lett. {\bf106}, 231802 (2011), derive an approximate formula for the UHE limit of $\sigma_{\nu N}(s)$ in a class of models that includes our own and assert that they are led ``to the important observation that $\sigma_{BBT}^{\nu N} \propto \ln^3s$, which manifestly violates the Froissart bound \cite{froissart} in contrast to what is stated in Refs. [6--8]'', the latter reference being to our work and the $\sigma_{BBT} ^{\nu N}$ to the cross sections we reported there. We here correct their erroneous implication that $\sigma_{\nu N}(s)$ should satisfy the Froissart bound and their mistaken assertion that we state that $\sigma_{BBT}^{\nu N}$ satisfies it.
\end{abstract}

There is long-standing interest in the behavior of the ultrahigh energy (UHE) neutrino-nucleon total cross section, $\sigma_{\nu N}(s)$, where $s\equiv (p_{\nu}+p_N)^2=2mE_\nu +m^2\approx 2mE_\nu$ is the  four-momentum squared of the neutrino-isobaric nucleon system, where $m$ is the proton mass and $N$ the isobaric nucleon.  The authors of a recent Physical Review Letters paper, ``Ultrahigh-Energy Neutrino-Nucleon Deep-Inelastic Scattering and the Froissart Bound",  \cite{ikk}, derive an approximate formula for the UHE limit of $\sigma_{\nu N}(s)$ in a class of models that includes our own and assert that they are led ``to the important observation that $\sigma_{BBT}^{\nu N} \propto \ln^3s$, which manifestly violates the Froissart bound \cite{froissart} in contrast to what is stated in Refs. [6-8]'', with {\em their Ref. [8]} being  {\em our work}---see our Ref.  \cite {bhm}---and the $\sigma_{BBT} ^{\nu N}$ being the cross sections \cite{bbt,bbmt} we reported in Ref.  \cite{bhm}.

Their above-quoted implication that $\sigma_{\nu N}(s)$ should satisfy the Froissart bound and their assertion about what is stated in our work are erroneous. In this commentary we cannot emphasize too strongly that we never implied, let alone \emph{stated}, that our calculation of the weak cross section $\sigma_{\nu N}(s)$ obeys the Froissart bound. It is simply \emph {not expected} to  satisfy $\sigma_{\nu N}(s) \leq \ln^2s$ \cite{froissart}; it is a first order perturbation expansion in the electroweak coupling constant $G_F$.  What we {\em do} say is that  the Froissart bound may apply to $\gamma^*p,\ W^{*\pm}N,\ Z_0^{*}N$ interactions \cite{bbt, bbmt,bhm}, which we treat on the same footing as \emph{strong} interactions. This is in the spirit of vector meson dominance, where  the electroweak currents communicate with hadronic systems through hadronic vector mesons \cite{schildknecht}. 

 Our high quality fit to \emph{all} of the small-$x$ HERA I data \cite{HERA} with a handful of parameters lends solid support to this hypothesis. In this case the deep inelastic scattering (DIS) reaction is $e+p\rightarrow e+X$, and the role of the electron is to supply the virtual $\gamma^*$ which, via vector dominance, interacts strongly with the proton. There is no reason to believe that $\sigma_{ep}(s)$, a quantity calculated only to leading order in $\alpha$, the electromagnetic coupling constant, is subjected to any bound.  Similarly, there is \emph{no} reason to expect that $\sigma_{\nu N}(s)$, the neutrino-isobaric nucleon total cross section---a quantity calculated in only {\em leading order} in electroweak interactions and all orders in strong interactions---is constrained by the Froissart bound. 

To make the distinction crystal clear between the interactions that might or might not be subjected to the Froissart bound, let us define the appropriate squared center of mass energy $s^* = (q+ p_N)^2$, where $q$ the off-shell vector boson 4-momentum. This is the final state hadronic invariant mass-squared (often denoted as $W^2$ in DIS literature); $s^*$ is the appropriate variable for the $\gamma^*p,\ W^{*\pm} N,\ Z_0^{*}N$ total strong interaction scattering cross sections. We reiterate that 
 the electron  simply furnishes the virtual photon $\gamma^*$ for DIS, whereas in the neutrino-nucleon interaction, the neutrino serves only to provide the $W^{*\pm}$ or the $Z_{0}^{*}$. 
Since for DIS, $s^*=Q^2(1-x)/x+m^2 \approx Q^2/x$ for small Bjorken $x$ and large, fixed virtuality $Q^2=-q^2$, saturating the Froissart bound of $\ln ^2(s^*)$ in $\gamma^*p$ scattering is equivalent to the statement that $F_2^{p}(x,Q^2)$ grows no faster than $\ln^2 x$. We emphasize that this is a statement \emph{only} about the structure function $F_2^{p}(x,Q^2)$; it does \emph{not} apply to the total $e p$ cross section's dependence on $s=(p_e+p_p)^2$.  Similarly, when the {\em same} proton structure function is used for off-shell, weak vector boson-$N$ scattering, no statement is made about $\sigma_{\nu N}(s)$ satisfying a Froissart bound in $s=(p_\nu +p_N)^2$.

The purpose of this comment is to make clear that our work in \cite{bbmt,bhm}  does not require or even hint at the possibility that the $\sigma_{\nu p}(s)$ cross section obeys the Froissart bound in $s$, using our model for $F_2(Q^2,x)$.  The Froissart  bound in $s^*$---used as a guide to the asymptotic behavior of $\sigma_{\gamma* N}(s^*) \sim F_2(Q^2,x)$---was its only application in our work. Thus, we conclude that the ``master formula'' of Ref. \cite{ikk}, their Eq.(10), applicable to models of $F_2(x,Q^2)$ including our own and that explicitly displays the models' asymptotic behaviors, {\em cannot} be used to determine whether or not $F_2$ has the ``correct'' asymptotic behavior. 

M. M. B. would like to thank Prof. A. Vainstein and Prof. L. Lipatov for valuable discussions.

\end{document}